\documentclass[%
 reprint,
 amsmath,amssymb,
 aps,
]{revtex4-2}

\usepackage{graphicx}
\usepackage{dcolumn}
\usepackage{bm}
\usepackage{array}
\usepackage{comment}
\usepackage{todonotes}
\usepackage[mathlines]{lineno}

\usepackage{glossaries}
\usepackage{tikz}
\newacronym{md}{MD}{molecular dynamics}

\begin{document}

\title{Interfacial Structure and Dynamics of Active–Passive Fluid Mixtures}

\date{\today}
\author{Linnea Heitmeier}
\email{heitmeier@ruc.dk}
\affiliation{``Glass and Time'', IMFUFA, Department of Science and Environment, Roskilde University, Postbox 260, DK-4000 Roskilde, Denmark}
 \affiliation{Institute of Frontier Materials on Earth and in Space, German Aerospace Center, Cologne, Germany}

\begin{abstract}
We investigate the interface between an active and a passive fluid using molecular dynamics simulations of active Brownian particles in contact with a passive Lennard-Jones fluid. 
We compute the surface tension via two independent routes, pressure anisotropy and capillary-wave analysis, and find that they disagree systematically once activity is switched on. Static density and orientation profiles reveal three distinct activity regimes, with particles preferentially oriented toward the interface. Going beyond these static properties, we characterize the interface dynamics: a threshold activity above which mixing becomes irreversible, coinciding with the onset of the mixing regime, and a transient overshoot in interfacial width before relaxation to steady state.

\begin{description}
\item[Keywords]
 Active Brownian particles, surface tension, molecular dynamics simulations
\end{description}
\end{abstract}

\newcolumntype{M}{>{$}c<{$}}
\glsresetall

\flushbottom
\maketitle
\thispagestyle{empty}
\setlength{\parindent}{0pt}
\section{Introduction}
Active particles are a prominent class of intrinsically out-of-equilibrium
systems, with realizations ranging from synthetic microswimmers to bacterial
suspensions and sperm. Because they convert stored or energy available from their surroundings into directed motion, they are promising for microrheology, drug delivery, and
micromachines \cite{patra2013intelligent, bechinger2016active, zottl2016emergent,
romanczuk2012active, wang2020active}. Most studies, however, address bulk
properties, whereas many applications depend just as strongly on the behavior
of active fluids near interfaces.

Consequently, the surface tension of active systems has attracted growing
attention, most commonly at free interfaces between an active fluid and vacuum
\cite{omar2020microscopic, lauersdorf2021phase, paliwal2017non}, in
motility-induced phase separation
\cite{cates2025active, fausti2021capillary, shi2020self, sun2025interfacial,
langford2024theory, burekovic2026active, fernandez2025dynamics, diaz2025activity},
in edge-current settings
\cite{van2016spatiotemporal, alsallom2026origin, adorjani2024phase,
zhou2025visual, caprini2025bubble, yang2020robust}, in wetting
\cite{grodzinski2026hydrodynamic, thiele2026active, turci2024partial,
chacon2022intrinsic, zhao2026wetting}, and via capillary rise
\cite{wysocki2020capillary}. These works report anomalous values, including
negative surface tensions for interfaces that nonetheless remain mechanically
stable \cite{bialke2015negative, Hermann2019Nonnegative, tjhung2018cluster}.
This paradox was recently resolved by Ref.~\onlinecite{li2025surface} for the
active--active interface between coexisting dense and dilute phases of active
Brownian particles, where densities are self-selected by activity.

Active--passive interfaces, by contrast, remain largely unexplored at the
particle-resolved level. Previous work has studied the mixing and structure of
active--passive fluids \cite{stenhammar2015activity, cugliandolo2017phase,
wysocki2016propagating}, propagating fronts \cite{xu2021morphologies}, and
kinetic roughening \cite{reichhardt2019reversibility}, while
Ref.~\onlinecite{adkins2022dynamics} described giant interfacial fluctuations,
active waves, and droplet emulsification in a continuum, flow-dominated regime.
None of these works quantifies the surface tension of the interface, tests
whether its thermodynamic and capillary-wave definitions, equivalent in
equilibrium, still agree under activity, or follows the interface after a
sudden change of activity.

Here, we close this gap using three-dimensional molecular dynamics simulations
of active Brownian particles in contact with a passive Lennard-Jones fluid.
We address two questions. Statically, we compare the thermodynamic and
capillary-wave routes to the surface tension as a function of activity.
Dynamically, we treat activity as a control parameter, analogous to a
quench in passive systems: we repeatedly switch it on and off, identify a
threshold activity above which mixing becomes irreversible on the timescale of
our simulations, and resolve the time-dependent interfacial width $\Delta(t)$
and position $z_0(t)$. This reveals a transient overshoot of the interfacial
width that is invisible in static analyses.

The active--passive setup is motivated twofold. First, fluid--fluid interfaces
are the relevant geometry in many biological and soft-matter applications,
unlike the fluid--vacuum interfaces considered in most of the literature above.
Second, a fluid--vacuum interface fixes the pressure on one side to zero,
whereas an active--passive interface lets us probe surface tension at finite,
tunable ambient pressure, a regime not systematically addressed for the
two-route comparison in active systems.

Beyond the stationary properties, the dynamics of active--passive interfaces
is of both fundamental and practical interest. In many realizations, activity
is not fixed but can be switched externally, e.g.\ via light or fuel
availability, so that the response to a sudden change of activity determines
how fast and how controllably such systems can be driven. Moreover, the
(ir)reversibility of activity-induced mixing probes memory and hysteresis,
which are absent for interfaces in equilibrium, and the relaxation of the
interface offers an independent test of whether equilibrium concepts such as
an effective surface tension remain meaningful out of equilibrium.

The paper is organized as follows. We first present the simulation method. We
then analyze static density profiles, which show three distinct activity
regimes, compare them quantitatively with capillary-wave theory, and examine
the particle orientation profiles. Finally, we turn to the interface dynamics,
covering the reversibility of activity-driven mixing and the time-resolved
relaxation of $\Delta(t)$ and $z_0(t)$ after a sudden change in activity.

\section{Methods}

\begin{figure}
    \centering
    \begin{tikzpicture}

        \node[anchor=south west, inner sep=0] (img) 
            {\includegraphics[width=0.3\linewidth]{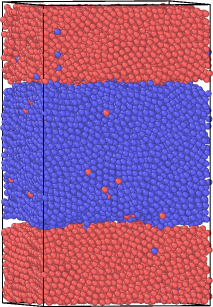}};

        \draw[->, thick] 
            ([xshift=-1.5cm]img.south west) 
            -- ([xshift=-1.5cm]img.north west)
            node[above] {$z$};

        \fill[blue] ([xshift=4.0cm,yshift=0.7cm]img.south west) circle (0.08cm);
        \node[right] at ([xshift=4.15cm,yshift=0.7cm]img.south west) {$A$ (active)};

        \fill[red] ([xshift=4.0cm,yshift=0.3cm]img.south west) circle (0.08cm);
        \node[right] at ([xshift=4.15cm,yshift=0.3cm]img.south west) {$B$ (passive)};

    \end{tikzpicture}

    \caption{Schematic view of the simulation setup. Blue particles correspond to the A-particles in this paper, the red ones correspond to the B-particles. }
    \label{fig:simulation_setup}
\end{figure}

We consider a binary system consisting of two particle species, $A$ and $B$. 
Along the $z$-direction, the system comprises two bulk regions (one containing the $A$-particles, the other the $B$-particles) separated by two interfaces normal to $z$. 
Periodic boundary conditions are applied in all three directions. 
A schematic view of the setup is shown in figure~\ref{fig:simulation_setup}. 

All particles interact via the Lennard-Jones potential
\begin{align}
    U(r_{ij}) = 4\epsilon_{\alpha \beta} \left[ \left(\frac{\sigma_{\alpha\beta}}{r_{ij}}\right)^{12} -  \left(\frac{\sigma_{\alpha \beta}}{r_{ij}}\right)^{6}\right] \ ,
\end{align}
where $r_{ij}$ is the distance between particles $i$ and $j$, and $\epsilon_{\alpha \beta}$ and $\sigma_{\alpha\beta}$ ($\alpha,\beta \in \{A,B\}$) set the energy and length scales, respectively. 
Any quantity can thus be expressed in units of $\sigma_{AA}$, $\epsilon_{AA}$, and the mass $m$; as is common practice, these units are omitted below. 
We use $\sigma_{AA} = \sigma_{BB} = \sigma_{AB}= 1.0$, $\epsilon_{AA}=\epsilon_{BB}=1.0$, and $\epsilon_{AB}=0.5$. 
The value of $\epsilon_{AB}$ controls whether the system mixes or demixes \cite{martinez2015liquid}.

The dynamics of particle $i$ are described by the standard model of active Brownian particles (ABPs) \cite{romanczuk2012active, cates2015motility}:
\begin{align}
    \frac{d \mathbf{r}^i}{dt} &= v_0^i  \hat{\mathbf{n}}^i - \mu^i \nabla U(\mathbf{r}^i) + \sqrt{2 D_T^i}\, \boldsymbol{\eta}_{\mathrm{trans}}(t) \ , \\
    \frac{d\theta^i}{dt} &= \sqrt{2 D_R^i}\, \eta_{\mathrm{rot}}(t) \ .
    \label{eq:equations_of_motion}
\end{align}
Here, $v_0^i$ is the self-propulsion velocity, $\mu^i$ the mobility, $D_T^i$ and $D_R^i$ the translational and rotational diffusion coefficients, and $\hat{\mathbf{n}}^i$ the orientation of the particle. Therefore, $n_z$ is the z-component of the orientation.
The noise terms $\eta$ are Gaussian white noise,
\begin{align}
    \langle \eta(t) \rangle = 0 \ , \qquad \langle \eta(t) \eta(t')\rangle = \delta(t - t') \ .
\end{align}
Each particle is thus an overdamped Brownian particle that can additionally propel itself. 
Because inertia is negligible, the velocity of a particle is set directly by the forces acting on it, weighted by its mobility $\mu^i$. 
The first equation contains three contributions to this velocity:
(i) the term $v_0^i \hat{\mathbf n}^i$ describes self-propulsion at constant speed along the instantaneous orientation $\hat{\mathbf n}^i = (\cos\theta^i, \sin\theta^i)$; it represents the conversion of energy from the environment into directed motion and drives the system out of equilibrium;
(ii) the term $-\mu^i \nabla U$ is the deterministic drift due to the interaction potential, e.g.\ steric repulsion between particles or an external potential;
(iii) the noise term $\sqrt{2D_T^i}\,\boldsymbol{\eta}_{\mathrm{trans}}$ accounts for thermal fluctuations of the surrounding solvent and leads to ordinary translational diffusion with coefficient $D_T^i$.

Rotational diffusion gradually randomizes the direction of self-propulsion, so that orientational correlations decay exponentially, $\langle \hat{\mathbf n}(t)\cdot\hat{\mathbf n}(0)\rangle = e^{-D_R^i t}$ in two dimensions. 
This defines a persistence time $\tau_R^i = 1/D_R^i$ and a persistence length $\ell_p^i = v_0^i/D_R^i$: on short times the particle moves ballistically, while on long times its motion is diffusive with an enhanced effective diffusion coefficient $D_{\mathrm{eff}}^i = D_T^i + (v_0^i)^2/(2D_R^i)$. 
For $v_0^i = 0$, self-propulsion vanishes and the model reduces to a passive Brownian particle, which relaxes to thermal equilibrium. 
The two species can differ in activity, mobility, and diffusivities, which allows us to describe mixtures of active and passive particles, as well as particles with different levels of activity.

The simulations were performed with the open-source, GPU-accelerated molecular dynamics package \texttt{gamdpy} \cite{gamdpy}. 
The system was initialized on a face-centered-cubic lattice with $15\times 15 \times 30$ unit cells at a target number density $\rho=0.8$. 
This gives a lattice constant $a=(4/\rho)^{1/3}\approx 1.71$, box dimensions $L_x=L_y=15a\approx25.6$ and $L_z=30a\approx51.3$, and $N=27000$ particles in total. 
Particles in the top half of the box ($z > L_z/2$) were assigned to species $A$, those in the bottom half to species $B$.

Unless stated otherwise, we use $D_T^A = D_T^B =0.005$, $D_R^A = D_R^B =1.0$, and $\mu^A = \mu^B = 0.05$. 
The equations of motion were integrated with a timestep $dt=0.005$. 
After an equilibration of duration $t=10^3$, we performed production runs of duration $t=10^7$. 
All reported density profiles are averages over $5$ independent initial configurations. 
For the study of the system's dynamics in section~\ref{sec:reversibility}, each phase had a duration of $t=512$.

During the production runs, we record the density profiles of both particle species as well as the height-resolved profiles of the particle orientations.

We calculate the surface tension from the pressure anisotropy of the system,
\begin{align}
    \hat \sigma = \int dz \left[ p_z(z) - \frac{1}{2}\big(p_x(z) + p_y(z)\big) \right] \ ,
    \label{eq:pressureanisotropy}
\end{align}
where $p_x$, $p_y$, and $p_z$ are the pressure components in the respective directions, determined from the virial (see \cite{gamdpy_stresssaver} for details).

To compare with capillary-wave theory \cite{buff1965interfacial}, we fit the density profiles to a $\tanh$ function,
\begin{align}
    \rho_{\mathrm{fit}}(z) = \frac{\rho_1 + \rho_2}{2} + 
     \frac{\rho_1 - \rho_2}{2} \tanh\left( \frac{z-z_0}{\Delta}\right) \ .
    \label{eq:tanh}
\end{align}
We call $\Delta$ the width of the density profile and $z_0$ the position of the interface. Note that due to symmetry reasons, we continue an analysis of only one of the surfaces. 
\section{Results}
\subsection{Static results}
\subsubsection{Density profiles}
\label{sec:density_profiles}
\begin{figure}
    \centering
    \includegraphics[width=\linewidth]{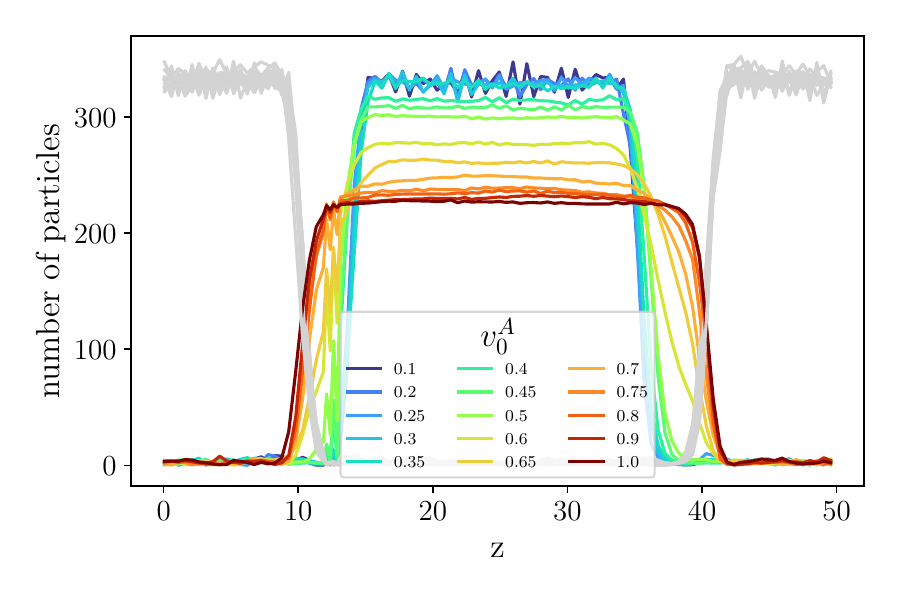}
    \caption{Density profiles of the active (colored) and passive (gray) particles for $D_T^A = 0.005$.  }
    \label{fig:density_profiles}
\end{figure}

In Fig.~\ref{fig:density_profiles} we
show the resulting density profiles of the active particles for
different self-propulsion velocities $v_0$, at a fixed translational
diffusion coefficient $D_T^A = 0.005$. It is convenient to express
the activity in terms of a translational P\'eclet number,
\begin{equation}
    Pe_t = \frac{v_0 \sigma_{AA}}{D_T^A},
    \label{eq:peclet}
\end{equation}
which quantifies the self-propulsion velocity relative to thermal
diffusion.

Two observations stand out immediately. First, the density profiles
of the passive B-particles remain essentially unchanged across the
entire range of activities studied. This is expected: activity
enters the particle dynamics of Eq.\eqref{eq:equations_of_motion} only through the
self-propulsion term $v_0^i \hat n^i$, which is identically zero for
the passive species; the B-particles are affected by the activity of
the A-particles only indirectly, through pair interactions localized
at the interface, and this local perturbation is too weak
to alter the passive bulk density profile appreciably. Second, and in
contrast, the activity of the A-particles has a pronounced effect on
their own density profile, and we can identify three qualitatively
distinct regimes.

For very small activities, $v_0 < 0.35$, corresponding to
$Pe_t < 70$, the density profile of the active particles remains
nearly indistinguishable from the passive case: thermal diffusion
dominates over self-propulsion, and the interface behaves
essentially as in an equilibrium binary fluid. For intermediate
activities, $0.35 < v_0 < 0.6$ (corresponding to $70<Pe_{t} <120$) the density profile progressively
broadens as $v_0$ increases, signaling that the active particles
increasingly accumulate against and press into the interfacial
region. Finally, for large activities, $v_0 > 0.6$,
the density profile broadens so much that active and passive
particles coexist within a particular region; we refer to
this as a \emph{mixing regime}. Notably, even in this regime the
density profile of the passive particles remains unchanged, which
shows that the mixing is asymmetric: active particles invade the
passive region, but the reverse process that passive particles
invading the active region does not occur to a comparable degree.

The bulk plateau value of the active-particle density profile,
$\rho_{\rm bulk} \approx 330$ (in units of particle number per bin),
is consistent with the nominal number density $\rho_{\rm total}=0.8$ and the box cross-section
$L_x \times L_y$, confirming that the systems remain well equilibrated
away from the interface irrespective of activity. As we show in the
following section, the width of the interfacial region continues to
grow with $D_T^A$ even at fixed $v_0$, and the activity at which the
crossover into the mixing regime occurs shifts to smaller $v_0$ as
$D_T^A$ increases which is consistent with the scaling suggested by
$Pe_t$ in Eq.~(\ref{eq:peclet}).


\begin{figure}
    \centering
    \includegraphics[width=\linewidth]{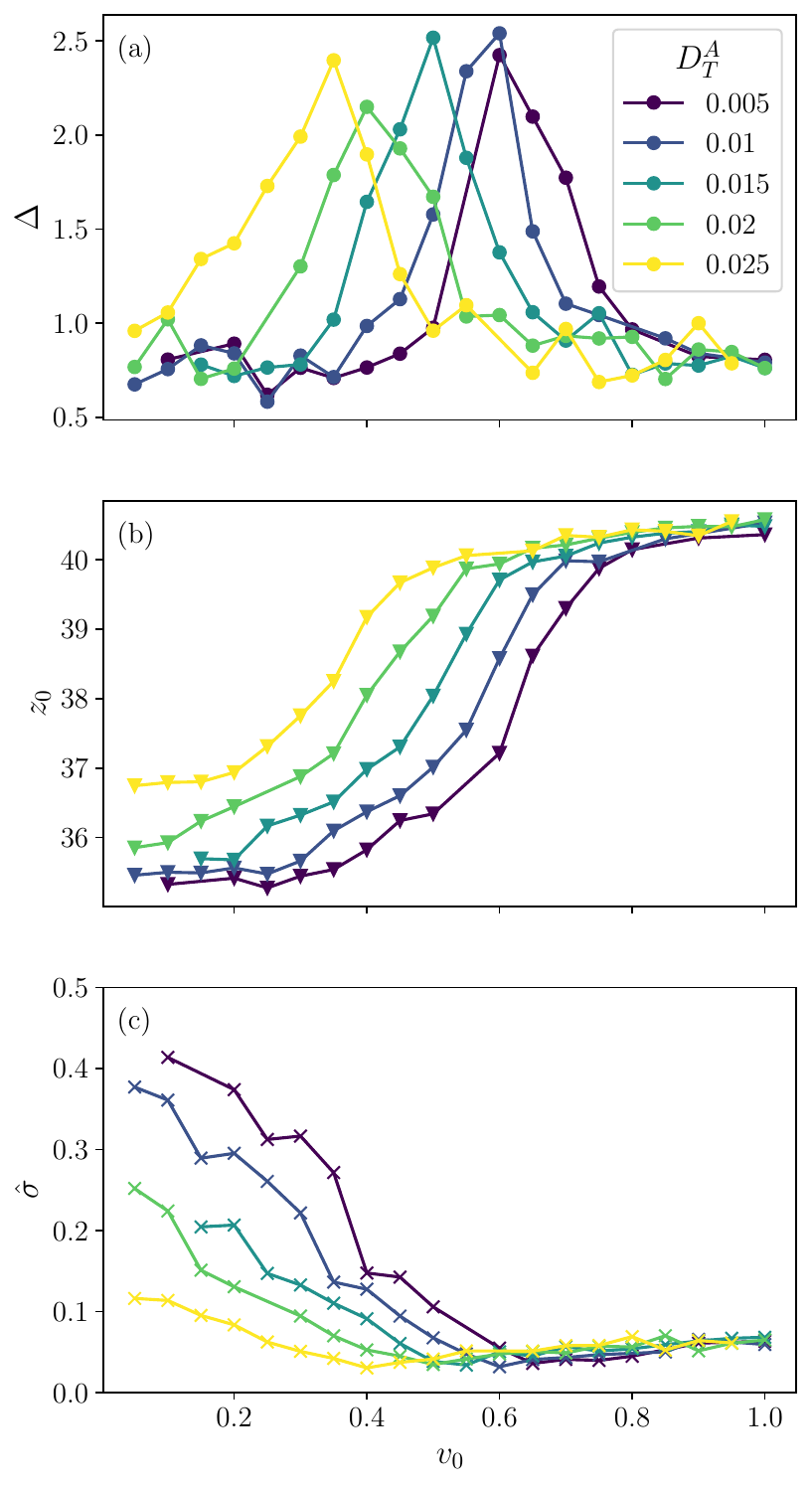}
    \caption{Density width, position and surface tension against activity for different thermal diffusion coefficients. }
    \label{fig:density_profiles_analysis}
\end{figure}

Figure~\ref{fig:density_profiles_analysis} presents a systematic analysis of
the density profiles and the surface tension as functions of
activity. For various combinations of $D_T^A$ and $v_0$, we fitted
the density profiles with Eq.~\eqref{eq:tanh} to extract the
interfacial width $\Delta$ and position $z_0$.

Panel (a) shows the width of the density profile as a function of
$v_0$ for different values of $D_T^A$. The behavior is strikingly
non-monotonic. For small $v_0$ and $D_T^A$, the width remains
essentially constant, consistent with the near-equilibrium regime
identified in Sec.~\ref{sec:density_profiles}, where the density
profile is barely perturbed by activity. As $D_T^A$ increases, this
constant regime progressively shrinks and eventually disappears. For
larger activities, the width first grows, reaches a pronounced peak,
and then decreases again. Notably, the location of this peak shifts
to smaller values of $v_0$ as $D_T^A$ increases, in line with the
scaling suggested by the P\'eclet number $Pe_t = v_0\sigma/D_T^A$
introduced above.

This non-monotonic behavior is itself a fingerprint of activity at a
finite-pressure interface. At a free active--vacuum interface, the
width of the density profile is expected to grow monotonically with
activity until the interface eventually becomes mechanically
unstable. In our simulations, by contrast, the active particles
cannot spread indefinitely: once they reach the passive fluid, the
finite density and pressure of the passive phase set an effective
upper bound on how far the interface can broaden, so that $\Delta$
saturates and subsequently decreases again as the system enters the
mixing regime. Consistent with this picture, the active--passive
interface in our simulations remains mechanically stable up to
substantially higher activities than what is reported for
active--vacuum interfaces, where instability sets in far earlier \cite{smrek2018interfacial, wysocki2016propagating}.

Panel (b) shows the corresponding position of the interface, $z_0$,
which reveals the same three regimes identified above: for very
small and for very large activities, $z_0$ is essentially
independent of $v_0$, while in the intermediate regime the interface
position shifts systematically toward the passive-particle side.
This shift indicates that, before the onset of full mixing, the
active fluid is already displacing the passive fluid and effectively
gaining volume at its expense.

Finally, panel (c) shows the surface tension $\hat\sigma$, obtained
independently from the pressure anisotropy, Eq.~\eqref{eq:pressureanisotropy}.
Here the trend differs qualitatively from that of $\Delta$: $\hat\sigma$
is maximal at vanishing activity, decreases monotonically with
$v_0$, and saturates at a small but non-zero plateau value of
$\hat\sigma \approx 0.05$--$0.07$ once the mixing regime is reached.
Remarkably, this plateau value is essentially independent of both
$v_0$ and $D_T^A$, suggesting that it reflects a generic property of
the fully mixed active--passive interface rather than a
activity-dependent quantity. The qualitative disagreement between
the non-monotonic behavior of $\Delta$ and the monotonic decay of
$\hat\sigma$ already indicates that density width and surface
tension are not simply proportional to one another once activity is
present; we examine this discrepancy quantitatively in the next
section.

\subsubsection{Comparison to capillary-wave theory}

The surface tension extracted from molecular dynamics simulations
can, in principle, be obtained via two independent routes. The first
is the thermodynamic, pressure-anisotropy definition of
Eq.~\eqref{eq:pressureanisotropy}. The second follows from
capillary-wave theory, which predicts that thermally excited
fluctuations of the interface broaden the density profile according
to \cite{buff1965interfacial}
\begin{align}
    \Delta^2 = \Delta_0^2 + \frac{k_B T}{2 \pi \hat\sigma} \log(L/B_0)
    := A + B\,\hat\sigma^{-1},
    \label{eq:cwt}
\end{align}
where $\Delta_0$ is an intrinsic offset width, $L$ is the lateral
size of the simulation box, and $B_0$ is a microscopic length scale.
Equation~\eqref{eq:cwt} predicts a linear relation between $\Delta^2$
and the inverse surface tension $\hat\sigma^{-1}$, with a slope set
by temperature alone. This is a relation that is well established in
equilibrium fluid-fluid interfaces but has to our knowledge not
been systematically tested for active-passive interfaces.

\begin{figure}
    \centering
    \includegraphics[width=\linewidth]{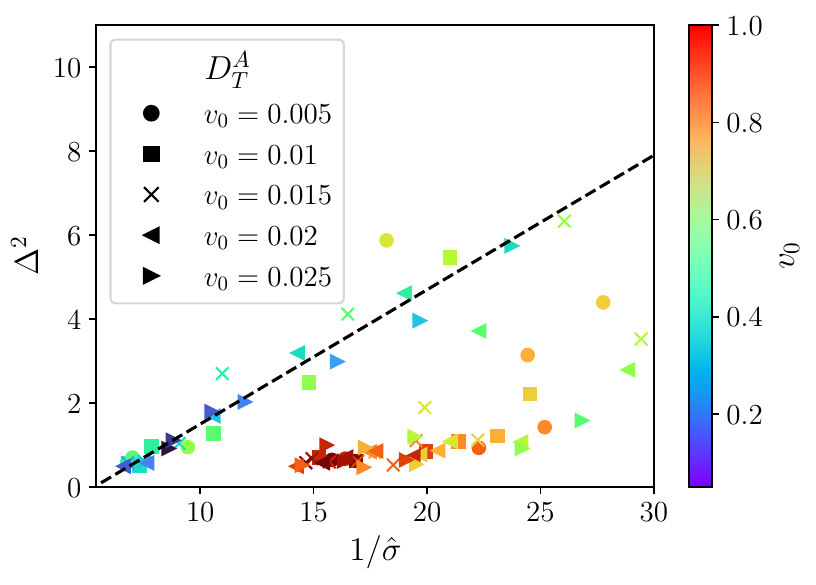}
    \caption{Test of the data against the prediction from capillary-wave theory.  }
    \label{fig:capillary}
\end{figure}

Figure~\ref{fig:capillary} tests this prediction against all data
presented in this paper, plotting $\Delta^2$ obtained from the
tanh-fits of Sec.~\ref{sec:density_profiles} against $\hat\sigma^{-1}$
obtained independently from the pressure anisotropy. For small
activities, the data indeed collapse onto the linear relation
predicted by Eq.~\eqref{eq:cwt}, confirming that capillary-wave
theory correctly captures the interfacial fluctuations of the
active-passive system in the near-equilibrium regime. As activity
increases, however, the data increasingly deviate from this linear
relation. This is consistent with the
qualitative disagreement between $\Delta(v_0)$ and $\hat\sigma(v_0)$
already noted in panels (a) and (c) of
Fig.~\ref{fig:density_profiles}: once the density profile broadens
due to genuine mixing of the two species, rather than due to
thermally excited capillary fluctuations alone, $\Delta$ ceases to be
a faithful proxy for $\hat\sigma$, and the two routes to the surface
tension no longer coincide.

This breakdown highlights a more fundamental ambiguity that is
specific to active systems: whereas the mechanical (pressure-based)
and fluctuation-based (capillary-wave) definitions of surface tension
are equivalent in equilibrium, this equivalence relies on the
assumption that pressure is a well-defined, isotropic bulk quantity
away from the interface. Note that this is an assumption that is itself
non-trivial for active matter, where swim pressure and interparticle
stresses need not combine into a conventional thermodynamic pressure
\cite{omar2020microscopic}. Our results suggest that this ambiguity
becomes appreciable precisely when the interface is broadened by
activity-driven mixing rather than by thermal capillary waves,
offering a concrete diagnostic for when the two definitions of
surface tension can be expected to agree.

\subsubsection{Orientations of the particles}

\begin{figure}
    \centering
    \includegraphics[width=\linewidth]{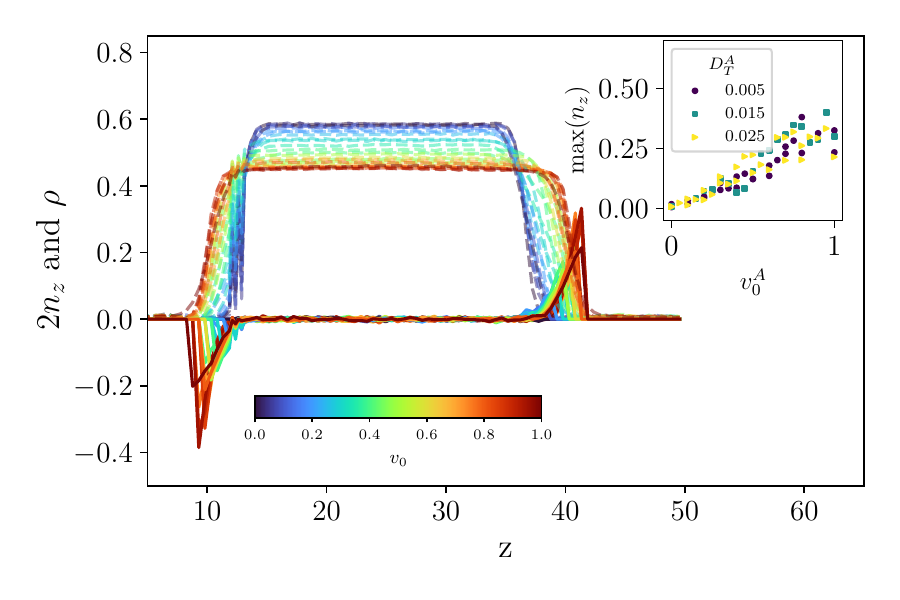}
    \caption{Height resolved orientation of the particles $n_z$ (full line) and density profile $\rho$ (dashed line) for different self-propelling velocities $v_0$ and constant $D_T^A=0.025$. The inset shows the maximum of the orientation profile in relation to $v_0$.  }
    \label{fig:orientation}
\end{figure}

Finally, in Fig.~\ref{fig:orientation} we examine one of the defining
properties of active matter that has no equilibrium counterpart: the
orientation of the self-propelled particles. The height-resolved
orientation profile reveals two distinct features. In the bulk
region, far from either interface, the mean orientation
$\langle n_z \rangle$ fluctuates around zero, as expected: in the
absence of any symmetry-breaking field, the swimming direction is
uniformly distributed and no net orientation should persist on
average. In the interfacial region, by contrast, the orientation
develops a pronounced net component directed toward the interface,
with active particles preferentially oriented so as to swim into the
passive phase.

This interfacial alignment has a simple mechanistic origin. Away
from the interface, an active particle's orientation evolves freely
via rotational diffusion, Eq. \eqref{eq:equations_of_motion}, unconstrained by its neighbors.
Close to the interface, however, particles oriented away from the
interface are pushed back into the bulk by steric repulsion from
their neighbors, while particles oriented toward the interface
continue to propel themselves forward unimpeded. This asymmetry
effectively biases the local population of orientations toward the
interface, even though the underlying rotational dynamics, in
Eq. \eqref{eq:equations_of_motion}, remains isotropic. The same mechanism is responsible for
particle accumulation at any confining or repulsive boundary in
active systems~\cite{elgeti2016microswimmers}, and here it operates locally at the
active-passive interface itself.

The inset of Fig.~\ref{fig:orientation} shows the maximal value of
this orientation, $\max(\langle n_z\rangle)$, as a function of $v_0$
for different values of $D_T^A$. Strikingly, the data for all values
of $D_T^A$ collapse onto a single curve, indicating that the degree
of interfacial alignment is controlled by the self-propulsion
velocity alone and is essentially insensitive to translational
diffusion. The collapsed data furthermore grow approximately
linearly with $v_0$, consistent with the interpretation above: since
the propulsive flux of particles into the interface scales directly
with $v_0$, while the counteracting steric restoring force is set
by the (activity-independent) interparticle interactions, the degree
of orientational bias inherits the linear dependence on $v_0$
directly.

\subsection{Dynamic results}
In this part, we investigate the dynamics of active-passive interfaces. This section is divided into 2 different parts: Firstly, we investigate the \emph{reversibility} of active-passive interfaces. After that, we investigate how the density profile at the interface develops. 

\subsubsection{Reversibility Threshold}
\label{sec:reversibility}
\begin{figure}[t]
    \centering
    \includegraphics[width=\linewidth]{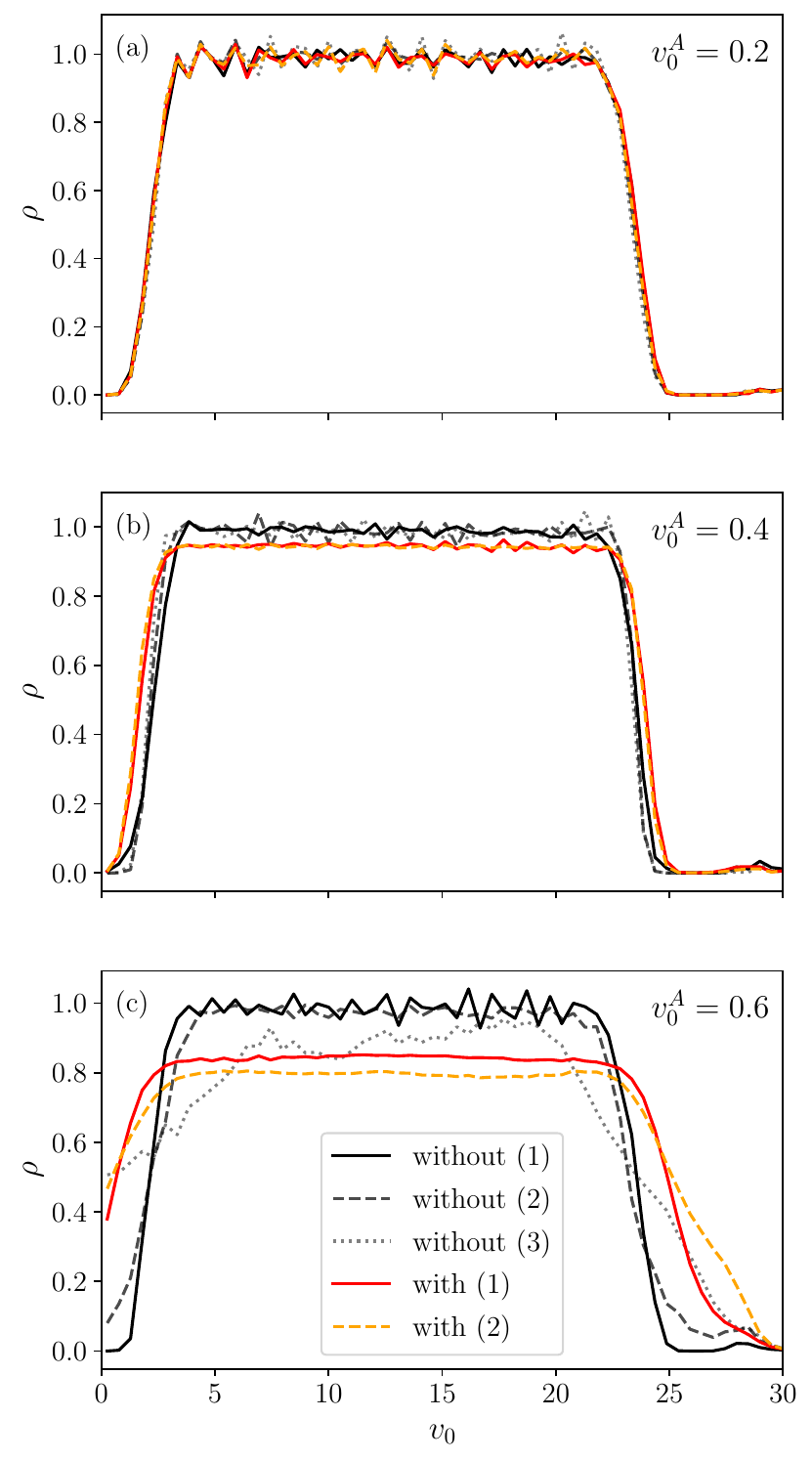}
    \caption{Density profiles for simulations where the activity was switched on and off ($D_T^A = 0.005$). One can see a transition between a reversible regime and a non-reversible regime: While for small activities, the passive density profile can be recovered when the activity is switched off, this is not the case for large self-propelling velocities. }
    \label{fig:reversibility}
\end{figure}

To probe the reversibility of the active-passive interface, we
perform simulations in which the activity is repeatedly switched on
and off, tracking the density profile at each stage. The protocol is
illustrated in Fig.~\ref{fig:reversibility}: after equilibration,
the system is first held in a passive state, denoted
\textit{without~(0)}; activity is then switched on
(\textit{with~(1)}), subsequently switched off again
(\textit{without~(1)}), and so on. If the underlying dynamics were
fully reversible, the density profile recorded in every
\textit{without} stage should be identical, independent of how many
activity cycles precede it.

This is indeed what we observe for small activities, $v_0 < 0.4$:
the passive density profile is recovered essentially unchanged after
each activation-deactivation cycle. For larger activities, however,
the \textit{without}-profiles from successive stages no longer
coincide, indicating that the system fails to return to its original
state once activity has been switched off -- the interface has been
irreversibly altered.

This loss of reversibility coincides with the onset of mixing
identified from the steady-state density profiles in
Sec.~\ref{sec:density_profiles}: for low activities, where the
active and passive fluids remain demixed at steady state, the
process is fully reversible, whereas for larger $v_0$, where active
and passive particles coexist within an extended interfacial region,
the process becomes irreversible. This correspondence suggests that
irreversibility is a direct consequence of mixing: Once active
particles have penetrated sufficiently deep into the passive phase,
purely diffusive relaxation after deactivation is no longer
sufficient to re-establish the original demixed configuration on the
timescale of our simulations.
\subsubsection{Activity timescale}
\begin{figure}
    \centering
    \includegraphics[width=\linewidth]{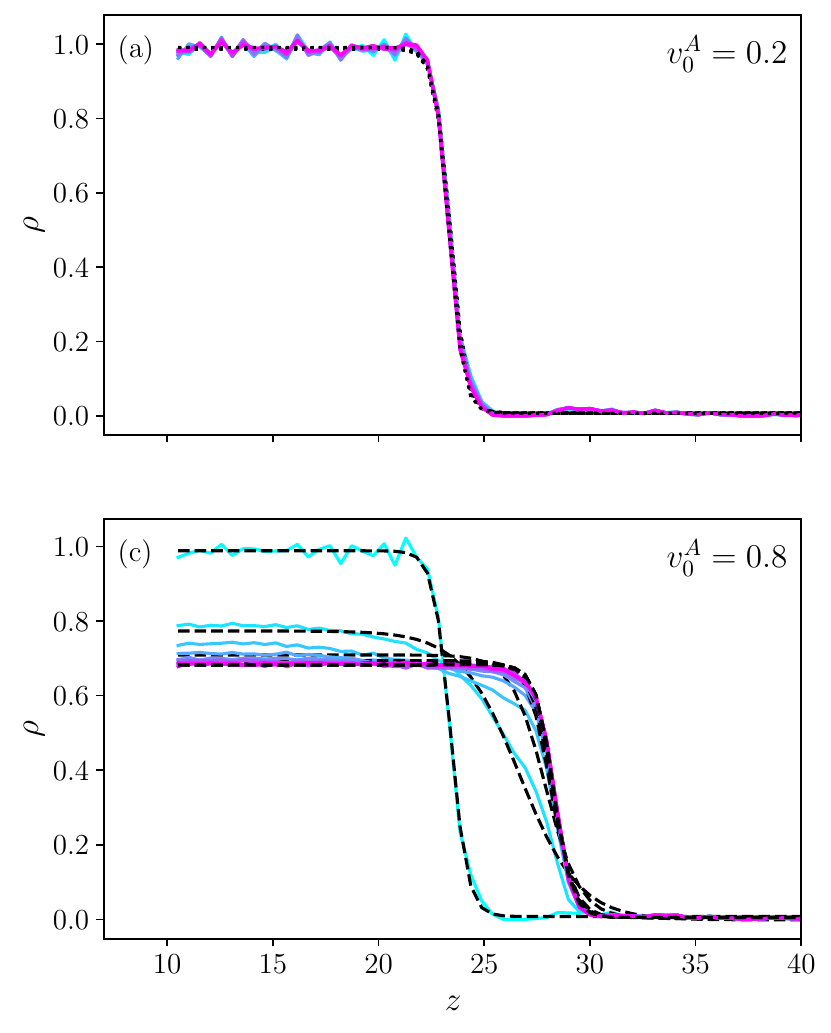}
    \caption{Density profiles for different activities at different times for a constant $D_T^A = 0.005$. Black dashed lines correspond to the fits with formula \eqref{eq:tanh}. }
    \label{fig:density_times}
\end{figure}

Having established that switching activity on and off within a
single configuration does not yield reversible statistics once
mixing has occurred, we instead characterize the transient dynamics
of interface formation using an ensemble approach: we prepare 1000
independent initial configurations, each equilibrated in the passive
state, and impose a fixed activity $v_0$ on all A-particles
simultaneously at $t=0$. This avoids the irreversibility artifacts
discussed in the previous section, at the cost of deviating from the
switch-on/switch-off protocol commonly used to probe relaxation in
passive systems \cite{heitmeier2025mode, knudsen2025hydrodynamic}.
Figure~\ref{fig:density_times} shows the resulting time evolution of
the density profile for several representative activities.

\begin{figure}
    \centering
    \includegraphics[width=\linewidth]{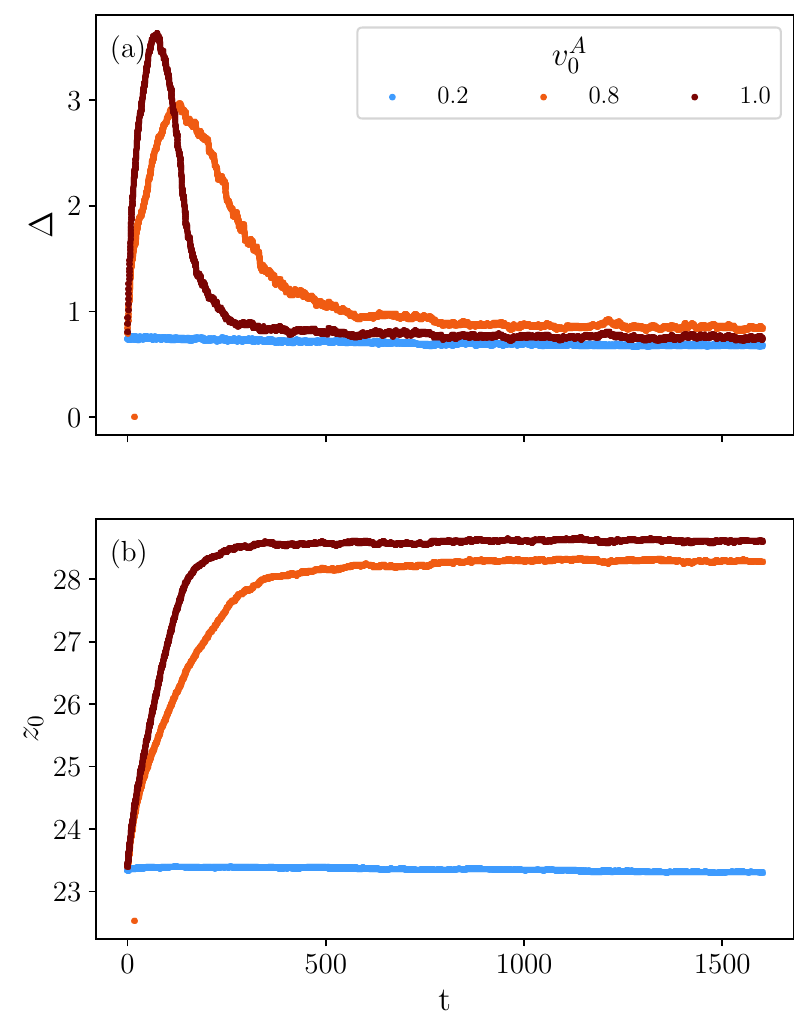}
    \caption{Time development of the width and the position of the density profile in time for a constant $D_T^A = 0.005$}
    \label{fig:timedevelopment}
\end{figure}

Figure~\ref{fig:timedevelopment} quantifies this transient by
tracking the fitted width $\Delta(t)$ and interface position
$z_0(t)$, obtained from Eq.~\eqref{eq:tanh}, as functions of time
for $D_T^A = 0.005$. Panel (a) shows a clear qualitative distinction
between low- and high-activity systems. For low activities, the
width remains essentially constant throughout the entire simulation,
consistent with the near-equilibrium, demixed steady state discussed
above. For higher activities, by contrast, $\Delta(t)$ first
overshoots sharply, reaching a pronounced transient peak, before
relaxing back down to a lower, quasi-stationary value. Both the
height and the timing of this peak depend strongly on activity: more
active systems overshoot earlier and more strongly, before
relaxing to their (still activity-dependent) steady-state width. This
transient overshoot suggests that the active particles initially
invade the passive phase faster than the interface can
mechanically respond, before the system settles into the broadened,
but stable, steady state characterized in
Sec.~\ref{sec:density_profiles}.

Panel (b) shows the corresponding evolution of the interface
position $z_0(t)$. All systems approach their steady-state position
monotonically, but the time required to do so depends systematically
on $v_0$: lower-activity systems reach their steady state
only after a substantially longer transient than higher-activity
systems. This trend is intuitive, since the rate at which particles
invade the passive region and hence the rate at which the
interface itself advances, is set directly by the self-propulsion
velocity.

\begin{figure}
    \centering
    \includegraphics[width=\linewidth]{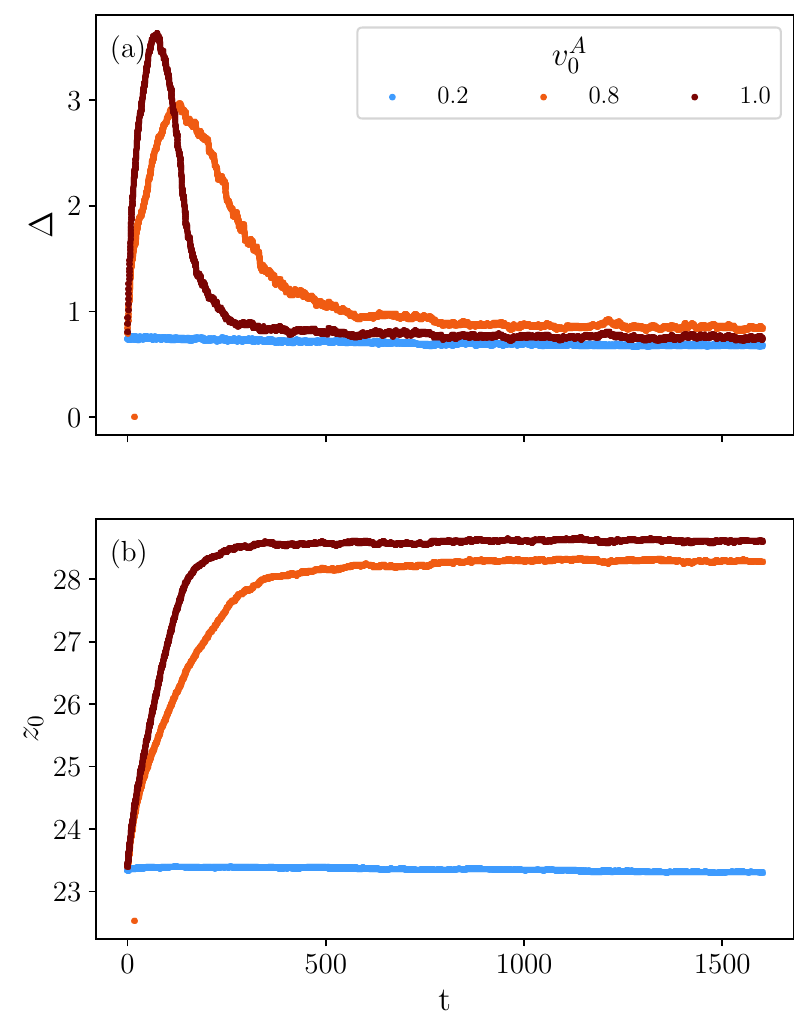}
    \caption{Time development of the width and the position of the density profile in time for a constant $D_T^A = 0.025$}
    \label{fig:timedevelopment_DT0.025}
\end{figure}

Figure~\ref{fig:timedevelopment_DT0.025} shows the same analysis for
a larger translational diffusion coefficient, $D_T^A = 0.025$. The
qualitative features, i.e. the transient overshoot in $\Delta$ and the
activity-dependent approach to steady state in $z_0$, are
reproduced, but occur at systematically shifted values of $v_0$: the
same qualitative behavior seen at a given $v_0$ for $D_T^A=0.005$ now
appears at correspondingly larger $v_0$ for $D_T^A=0.025$. This
shift is consistent with the P\'eclet-number scaling,
$Pe_t = v_0\sigma/D_T^A$, already identified in the static analysis
of Sec.~\ref{sec:density_profiles}, and confirms that the
same underlying control parameter governs both the static and the
dynamic interfacial behavior.


\section{Summary and outlook}

In this work, we investigated the interface between an active and a
passive fluid using 3-dimensional molecular dynamics simulations of active
Brownian particles in contact with a passive Lennard-Jones fluid.
Unlike the free active--vacuum interfaces studied previously, this
setup allowed us to probe the interfacial behavior of active matter
at finite ambient pressure. Our results contribute in two
respects: a comparison of two routes to the surface tension at finite
ambient pressure, and, to our knowledge for the first time, a
particle-resolved characterization of the interface dynamics
following a sudden change in activity.

Both the width and the position of the interfacial density profile
depend non-monotonically on activity, with three distinct regimes: a
near-equilibrium regime at low $Pe_t$ in which the interface is
essentially unperturbed, an intermediate regime in which the active
fluid progressively invades the passive phase, and a high-activity
mixing regime in which active and passive particles coexist over an
extended region. The surface tension obtained from the pressure
anisotropy, by contrast, decays monotonically with activity and
saturates at a small, non-zero plateau value that is independent of
both $v_0$ and $D_T^A$. This qualitative disagreement between the two
observables shows that the width of the density profile is not simply
proportional to the surface tension once activity is present, in
contrast to the equilibrium case. Testing this connection
quantitatively against capillary-wave theory, we found good agreement
in the near-equilibrium regime, but systematic deviations once the
system enters the mixing regime, indicating that the broadening of
the interface beyond a certain activity is driven by genuine mixing
rather than by thermally excited capillary fluctuations. We further
characterized the orientation of the active particles across the
interface, and found that particles preferentially orient toward the
interface, with the degree of alignment controlled by activity alone.

Going beyond these static properties, which have so far been the
focus of studies of active--passive interfaces, we investigated the
interfacial dynamics. The interface exhibits a threshold in activity
beyond which the mixing process becomes irreversible: activity can no
longer be switched off to recover the original, demixed state within
the accessible simulation time. This loss of reversibility roughly
coincides with the onset of the mixing regime identified in the
static analysis, suggesting that both phenomena share a common
origin. Tracking the time evolution of the density profile following
a sudden onset of activity, we observed a pronounced transient
overshoot in the interfacial width before the system relaxes to its
steady state, with the timing and magnitude of this overshoot
controlled by activity in a manner consistent with the P\'eclet-number
scaling identified in the static case. To our knowledge, neither the
reversibility threshold nor the time-resolved relaxation of the
interfacial width and position has previously been characterized for
active--passive fluid interfaces. 

Taken together, our results demonstrate that the surface tension of
an active-passive interface cannot, in general, be inferred from
capillary-wave theory alone, and that the two standard routes to the
surface tension, which coincide in equilibrium, need not agree
once the system is driven out of equilibrium by activity, even when
the interface remains mechanically stable throughout. This raises a
more fundamental question that our results do not resolve: how
pressure, and hence surface tension, should be defined in
non-equilibrium active systems in the first place, given that swim
pressure and interparticle stresses need not combine into a
conventional, isotropic bulk pressure far from the interface
\cite{omar2020microscopic, yu2025pressure, solon2015pressure, takatori2014swim}.

Several directions emerge naturally from this work. First, a
systematic comparison of alternative pressure definitions (e.g.,
mechanical versus virial routes) at the active-passive interface
would help clarify when the disagreement between thermodynamic and
capillary-wave surface tensions can be attributed to a genuine
breakdown of capillary-wave theory, as opposed to an ambiguity in the
definition of pressure itself. Second, the present dynamic analysis
was restricted to the interfacial width and position; a
time-resolved measurement of the surface tension itself during the
transient following the onset of activity would show whether the
overshoot in the width is mirrored in the thermodynamic surface
tension, and would extend the comparison of the two routes to the
time domain. Third, our results were obtained at a single interaction
asymmetry ($\epsilon_{AB}=0.5$); a systematic study of a variation
of $\epsilon_{AB}$ to tune the equilibrium miscibility of the two
species would establish how robust the observed regimes are, and
allow the transition into the mixing regime to be located more
precisely. Fourth, a systematic variation of the activity of the
second species from passive to as active as the first would show
whether the mixing threshold shifts once both species can invade each
other symmetrically.

Finally, given that mixed active-passive interfaces of the kind
studied here are relevant to biological systems in which motile and
non-motile constituents coexist, extending this analysis to more realistic interaction
potentials and confinement geometries would be a natural next step
toward connecting these results to biological soft matter.

\begin{acknowledgments}
The author acknowledges fruitful discussions with Jeppe Dyre, Sumeja Burekovi\'c and Søren Toxværd. 
This work was supported by the Independent Research Fund Denmark through grant no. 5281-00148B.

\end{acknowledgments}
\bibliography{keylatex2}

\end{document}